\documentclass{article}
\usepackage{ijcai19}

\usepackage{times}
\usepackage{soul}
\usepackage{url}
\usepackage[hidelinks]{hyperref}
\usepackage[utf8]{inputenc}
\usepackage[small]{caption}
\usepackage{subcaption}
\usepackage{graphicx}
\usepackage{xcolor}
\usepackage{amsmath}
\usepackage{booktabs}
\usepackage{algorithm}
\usepackage{algorithmic}
\title{LogicLens: Leveraging Semantic Code Graph to explore Multi Repository large systems}

\author{
Niko Usai$^1$\and
Dario Montagnini$^1$\and
Kristian Ilianov Iliev$^1$\And
Raffaele Camanzo$^1$
\affiliations
$^1$Sourcesense\\
\emails
\{niko.usai, dario.montagnini, kristian.iliev, raffaele.camanzo\}@sourcesense.com
}

\definecolor{todo_color}{RGB}{219, 48, 122}

\begin{document}

\maketitle

\begin{abstract}
Understanding large software systems is a challenging task, especially when code is distributed across multiple repositories and microservices. Developers often need to reason not only about the structure of the code, but also about its domain logic and runtime behaviors, which are typically implicit and scattered.

We introduce \textbf{LogicLens}, a reactive conversational agent that assists developers in exploring complex software systems through a semantic multi-repository graph. This graph is built in a preprocessing step by combining syntactic code analysis, via AST parsing and repository traversal, with semantic enrichment using Large Language Models (LLMs). The resulting graph captures both structural elements, such as files, classes, and functions, as well as functional abstractions like domain entities, operations, and workflows.

Once the graph is constructed, LogicLens enables developers to interact with it via natural language, dynamically retrieving relevant subgraphs and answering technical or functional queries. We present the architecture of the system, discuss emergent behaviors, and evaluate its effectiveness on real-world multi-repository scenarios.

We demonstrate emergent capabilities including impact analysis and symptom-based debugging that arise naturally from the semantic graph structure.

\end{abstract}

\section{Introduction}

Modern software systems are increasingly complex and distributed, often composed of multiple repositories, services, and technologies. For example, developers working on large-scale systems often struggle to answer questions such as: 
\textit{``Which services are involved in the processing of a user payment?"}, 
\textit{``Where is a given business rule implemented across repositories?"}, or 
\textit{``What components must be updated to safely modify a data schema?"} 
Answering these questions typically requires manual inspection of multiple codebases, documentation, and runtime configurations.
 Developers face significant challenges when trying to understand both the structural and functional aspects of such systems \cite{newman2015building,bass2015devops}. Traditional tools for code navigation and static analysis focus on isolated repositories and lack the semantic abstraction needed to reason about cross-cutting concerns or system-wide workflows \cite{koschke2002survey}.
The rise of \textit{Software Knowledge Graphs} has introduced promising techniques for representing codebases as graph structures, capturing relationships between code elements and supporting advanced queries \cite{abdelaziz2021toolkit,liu2025codexgraph}. 
Similarly, Large Language Models (LLMs) \cite{vaswani2023attention} have shown impressive capabilities in code summarization, semantic search, and conversational interfaces \cite{zhang2023survey,chang2024survey}. However, these models often lack explicit access to a system-level representation of the software, limiting their ability to answer high-level or cross-repository questions.

To address these challenges, we introduce \textbf{LogicLens}, a conversational agent designed to assist developers in navigating and understanding complex software systems. At the core of LogicLens is a semantic graph, which combines structural and functional information from across repositories. This graph is constructed in a preprocessing phase that merges syntactic analysis through AST parsing and repository traversal with semantic enrichment using Large Language Models (LLMs). The resulting graph represents not only the technical structure of the system, such as files, classes, and functions, but also higher-level functional abstractions like domain entities, operations, and workflows.
Once constructed, the graph serves as the foundation for interactive querying. LogicLens allows developers to interact with the system using natural language, dynamically retrieving relevant subgraphs and providing responses to both technical and domain-specific questions. By selecting tools reactively based on user input, the agent is able to provide answers that span multiple codebases and services.
This paper explores the architecture of LogicLens, highlighting the emergent behaviors it enables, and evaluating its effectiveness in real-world multi-repository software scenarios. We show how this approach supports developers to reason more efficiently about complex systems and navigate the implicit domain knowledge embedded in large codebases.

\section{Related Work}


Understanding and navigating codebases has been a longstanding challenge in software engineering \cite{fowler2024legacy}. Various tools have been developed to support semantic code search, summarization, and analysis, primarily targeting single repositories.

\textbf{Code Search and Analysis Tools.} 
Sourcegraph \cite{sourcegraph} offers advanced code navigation and search capabilities across repositories, but its semantic reasoning is limited and it primarily supports syntactic queries within individual repositories. 
CodeQL \cite{codeql} enables static analysis through a query language designed for security and quality checks, but it lacks semantic enrichment and cross-repository reasoning. 
LLM-based coding tools such as Claude Code \cite{anthropic2024claudecode} and systems built on OpenAI Codex \cite{chen2021evaluating} provide conversational interfaces for code understanding, generation, and refactoring, but they typically operate over limited context windows and remain focused on individual repositories or local code fragments.
Similarly, AI-powered assistants such as GitHub Copilot \cite{copilot} and Cursor \cite{cursor} leverage large language models to assist developers, yet they lack an explicit system-level representation of software and do not support deep cross-repository reasoning.

\textbf{Semantic Search and Summarization.} Approaches like CodeSearchNet \cite{husain2019codesearchnet} have advanced semantic code retrieval within individual projects by leveraging neural embeddings. Tools like Aroma \cite{lv2019aroma} provide semantic code search focused on snippet retrieval. Windsurf \cite{windsurf} introduces conversational interaction with codebases, yet remains limited to individual repositories.

\textbf{Software Knowledge Graphs and LLM Agents.} Recent approaches build software knowledge graphs to represent code elements and their relations for enhanced code comprehension \cite{zhang2021softwareKG,gupta2021neural}. Large Language Models have also been used to build conversational agents that reason over code, yet they typically lack explicit integration with multi-repository semantic graphs.

Our approach differs from existing tools and methods along two key dimensions. 
First, while prior code search and analysis systems operate primarily on syntactic representations or local semantic embeddings, LogicLens relies on an explicit semantic graph that captures both structural dependencies and domain-level abstractions. 
Second, unlike LLM-based assistants that reason over limited context windows, our approach provides the agent with a persistent, multi-repository system representation, enabling queries and reasoning that span repositories, services, and end-to-end workflows. 
This combination allows LogicLens to support system-level exploration and cross-cutting queries that are not addressed by existing single-repository or context-bounded approaches.

\section{Method}

\begin{figure*}[!t]
     \centering
     \begin{subfigure}[b]{0.49\textwidth}
         \centering
         \includegraphics[width=\textwidth]{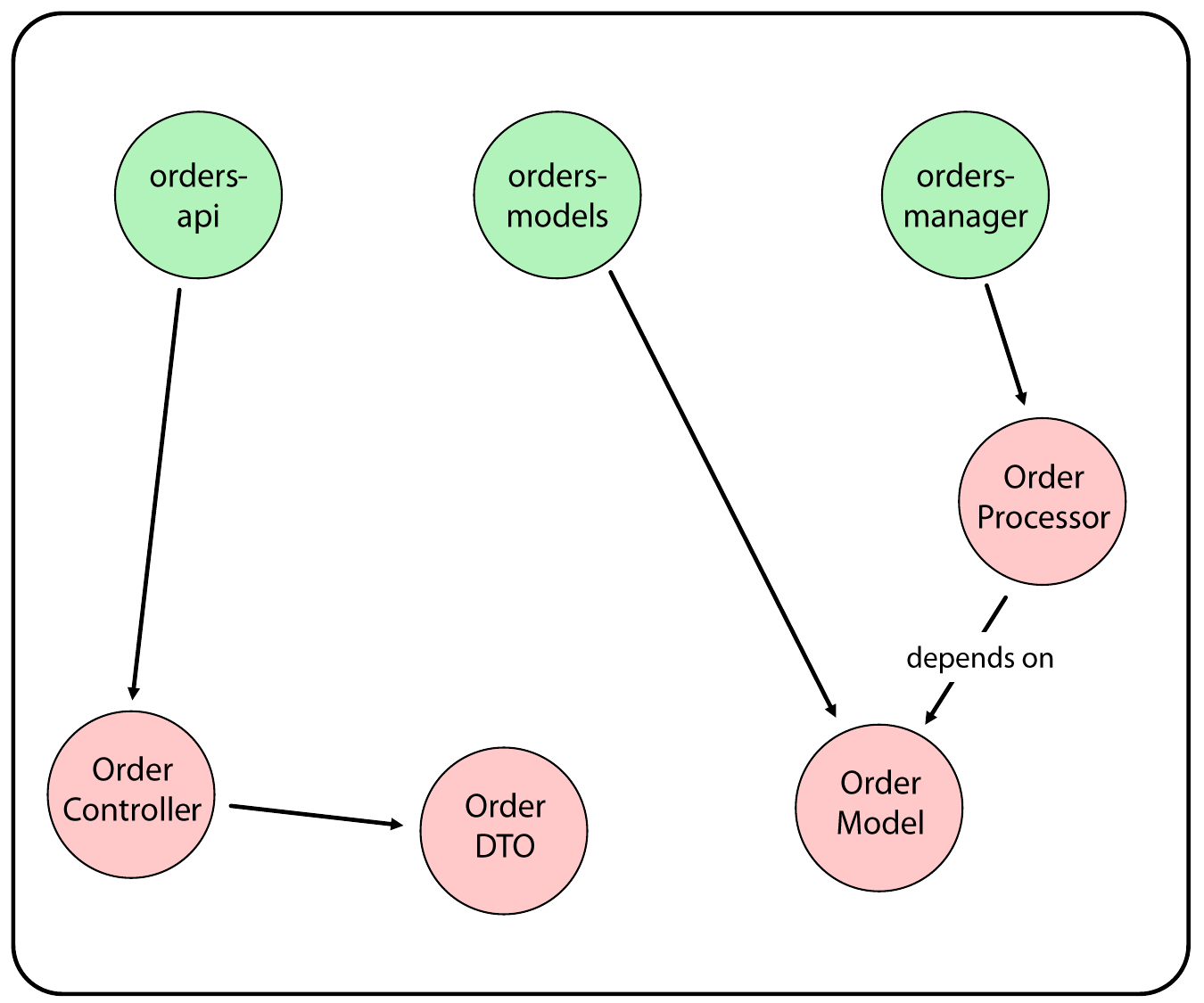}
         \caption{Structural Graph}
         \label{fig:structural_graph}
     \end{subfigure}
     \begin{subfigure}[b]{0.49\textwidth}
         \centering
         \includegraphics[width=\textwidth]{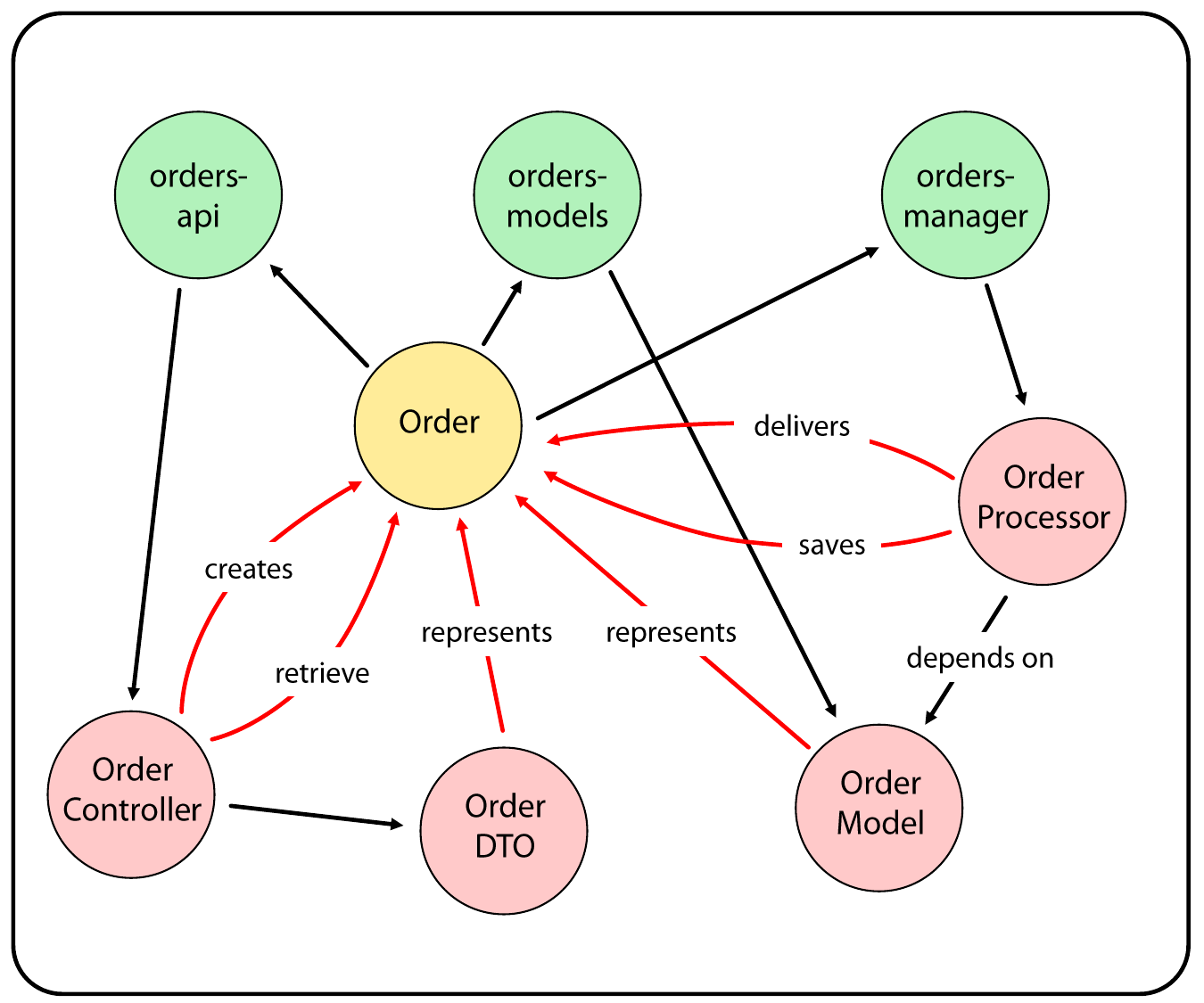}
         \caption{Semantic Graph}
         \label{fig:semantic_graph}
     \end{subfigure}
    \caption{Comparison between the structural and semantic graphs. 
    (a) Structural Graph: captures static relationships among code units across multiple repositories, including dependencies, method calls, and inheritance. 
    (b) Semantic Graph: enriches the structural graph with domain entities and functional relationships, connecting code units through the operations they perform on shared entities. 
    For example, a \texttt{Code} node representing \texttt{OrderProcessor} is linked to an \texttt{Entity} node \texttt{Order}, illustrating how multiple services collaborate on the same domain concept.}

\end{figure*}

Our objective is to develop a system that enables the exploration and troubleshooting of large-scale systems—whether microservices-based or monolithic—comprising multiple repositories, through a conversational interface.

This interface should allow users, even those not deeply familiar with the system, to ask questions such as:
\begin{itemize}
    \item What is the creation flow of resource X? (explore)
    \item Which projects are involved in the activation process of Y? (explore)
    \item The deletion of Z is failing; which projects should I inspect to identify the root cause? (troubleshoot)
    \item Endpoint Q returns a 500 error—what might be causing it? (troubleshoot)
\end{itemize}

We leverage a ReAct Agent~\cite{yao2023reactsynergizingreasoningacting} architecture combined with GraphRAG techniques to retrieve comprehensive context from a Knowledge Graph built solely from source code, in order to answer user queries.

Our approach consists of firstly constructing a Knowledge Graph from multiple repositories, and then employing a ReAct Agent to answer user queries exploiting this graph. To illustrate the approach, we introduce a running example centered around a generic Order Management system. This example will guide the explanation of the graph construction phase with more specific details.

The system is composed of three repositories:
\begin{itemize}
    \item \texttt{orders-api}: a REST API that asynchronously receives orders.
    \item \texttt{orders-models}: a shared library containing domain models.
    \item \texttt{orders-manager}: a consumer service responsible for order processing.
\end{itemize}

\subsection{Graph Construction}

The graph construction process follows a three-stage workflow. First, an algorithmic phase extracts structural information from the codebase to build a structural code graph. Second, a generative phase enriches the graph nodes with semantic descriptions. Finally, a concept extraction phase adds a functional dimension to the graph by identifying and representing domain-specific concepts.

\paragraph{Structural Graph.}
The first stage consists of extracting a structural representation of the system from the source code of each repository. The goal is to capture static relationships among code units, such as dependencies, inheritance, and invocations, within and across repositories.

For each code repository, we generate a structural subgraph composed of:
\begin{itemize}
    \item a \texttt{Project} node representing the repository;
    \item multiple \texttt{Code} nodes, each representing a relevant \textbf{code unit} which is a language-specific meaningful container, such as a class in Java or a function or struct in Golang.
    \item typed edges between \texttt{Code} nodes to indicate structural relationships (e.g., dependencies, method calls, inheritance).
\end{itemize}

All \texttt{Project} nodes are linked to a top-level \texttt{System} node, enabling queries that span across multiple repositories. This design also supports the modeling of shared libraries that connect otherwise independent codebases.

Figure~\ref{fig:structural_graph} shows an example of the resulting structural graph for a simplified Order Management system. In this example:
\begin{itemize}
    \item the \texttt{orders-api} project contains the \texttt{OrderController} and \texttt{OrderDTO} classes;
    \item the \texttt{orders-models} library defines a shared \texttt{OrderModel};
    \item the \texttt{orders-manager} service includes an \texttt{OrderProcessor} that depends on the \texttt{OrderModel}.
\end{itemize}
Edges between nodes capture static dependencies, such as class usage or field references, e.g., \texttt{OrderProcessor} \texttt{depends on} \texttt{OrderModel}.

To extract the \texttt{Code} nodes and their relationships, we parse the Abstract Syntax Tree (AST) of each source file using the Tree-sitter library  \cite{treesitter2024}. Language-specific queries are used to identify code units and extract relationships based on supported programming languages. In its current state, LogicLens supports Java, Python, Go, and TypeScript.
Considering a Java example, the \texttt{(Code)-[:DEPENDS\_ON]->(Code)} relationship is extracted from constructs such as \texttt{extends}, \texttt{implements}, imports, or field types.

Each \texttt{Code} node is assigned a unique identifier. The identifier strategy is language-specific; in Java, we use the fully qualified name (FQN) to ensure uniqueness across repositories. This makes it possible to resolve and unify references to shared types across projects, such as common data models.

At this stage, the structural graph is described by the following relationship schema:

\begin{verbatim}
(System)-[:CONTAINS]->(Project)
(Project)-[:CONTAINS]->(Code)
(Code)-[:DEPENDS_ON]->(Code)
(Code)-[:CALLS]->(Code)
(Code)-[:IMPLEMENTS]->(Code)
\end{verbatim}

This structural graph captures low-level static relationships between code components and repositories. However, it does not yet represent higher-level semantics or domain-specific concepts, which are addressed in the following phases.

\paragraph{Semantic Graph.}
In the second phase of the graph construction process, we move beyond purely structural information to enrich the graph with semantic context. We enrich the \texttt{Code}, \texttt{Project}, and \texttt{System} nodes with natural language descriptions generated via a hierarchical prompting strategy using Large Language Models (LLMs). These descriptions enable semantic indexing and facilitate retrieval of relevant nodes during query resolution.

For each \texttt{Code} node, a concise summary of three sentences is generated from the source code context using fast, cost-effective LLMs such as GPT-5 mini or Claude 4.5 Haiku. These summaries aim to explain the purpose, structure, and behavior of individual functions or classes.

\texttt{Project} node descriptions are generated by aggregating the descriptions of the underlying \texttt{Code} nodes within the same repository. These summaries emphasize the core responsibilities and functional role of the project in the overall system architecture—such as whether it acts as an API interface, a data model library, or a background worker.

At the highest level, the \texttt{System} node description synthesizes all project summaries to provide a unified, high-level overview of the system's functionality and purpose. Descriptions for both \texttt{Project} and \texttt{System} nodes are generated using more capable LLMs such as GPT-5 or Claude 4.5, which are better suited for tasks requiring abstraction and synthesis of multiple inputs. All this information captures the static relationships between code components and repositories, but it does not provide insight into the functional behavior of the system or how multiple projects collaborate to implement domain workflows. 
This limitation motivates the next phase, where we enrich the structural graph with semantic information to capture high-level concepts, operations, and domain entities.

\paragraph{Semantic Graph Enhancement.}
To capture the functional and domain-specific aspects of the system, we augment the structural graph with semantic information. They include natural language descriptions of code units and projects, domain-specific entities, and functional relationships. This enrichment is important because it allows to understand not only the static structure of the system but also its behavior, workflows, and how multiple projects collaborate on the same domain concepts.

Building upon the \texttt{Code} nodes, we construct context by generating a new type of nodes called \texttt{Entity} nodes. An \texttt{Entity} is a concept with a strong semantic meaning, representing something identifiable and significant, and capturing domain-specific concepts related to the system. The introduction of \texttt{Entity} helps in establishing relationships between projects at the semantic level of functionality, transcending the purely structural aspects of the code. These nodes are generated using specialized prompts and deep LLMs. An example of the entity-enhanced graph is shown in Figure \ref{fig:semantic_graph}.

Unlike the static relationships produced in the structural phase, the edges connecting \texttt{Code} nodes to \texttt{Entity} nodes have dynamic labels representing verbs or actions. In fact, different code units may perform different operations on the same entity (e.g., \texttt{CREATE}, \texttt{PRODUCE}, \texttt{CONFIGURE}). As regards \texttt{Code} nodes containing model objects, entities, or DTOs, the primary relationship encoded is \texttt{(Code)-[:REPRESENTS]->(Entity)}. Project and entity nodes are then linked by a \texttt{RELATES\_TO} edge. The relationships created at this step have the following schema:

\begin{verbatim}
(Code)-[:CREATE]->(Entity)
(Code)-[:PRODUCE]->(Entity)
(Code)-[:CONFIGURE]->(Entity)
(Code)-[...]->(Entity)
(Code)-[:REPRESENTS]->(Entity)
(Entity)-[:RELATES_TO]->(Project)
\end{verbatim}

Previously, a purely structural graph only captured low-level dependencies, making it difficult to trace workflows or understand how multiple projects collaborate on the same domain concept. By introducing \texttt{Entity} nodes, code units and projects are connected through the entities they operate on, creating functional bridges across repositories. This results in a more connected and semantically meaningful graph, where domain entities serve as hubs linking projects and code units, enabling queries that follow workflows and operations across multiple projects.

\subsection{Agent Architecture}

\begin{figure*}[h]
  \centering
  \includegraphics[width=0.98\textwidth]{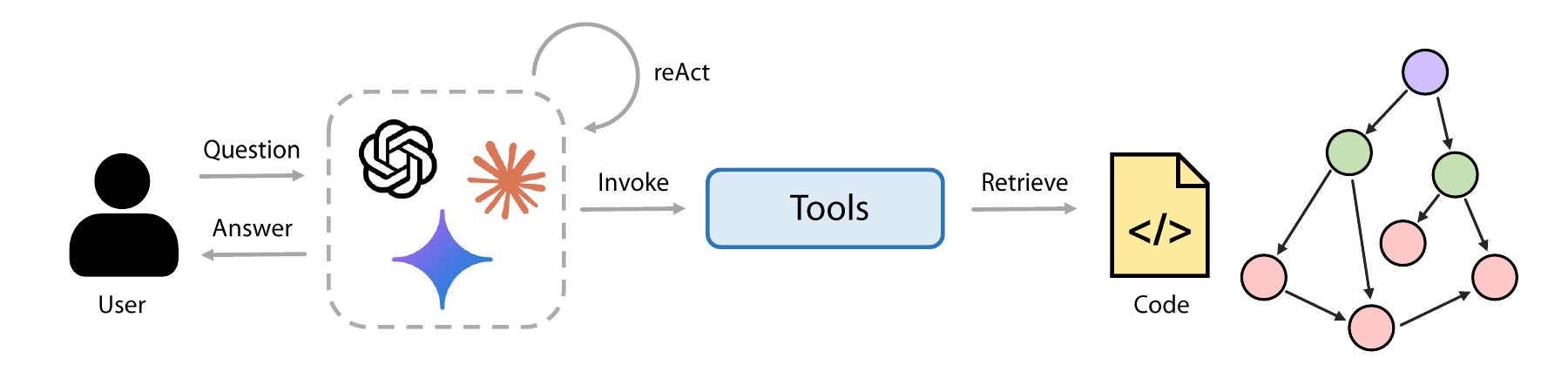}
  \caption{ReAct Agent}
  \label{fig:react_agent}
\end{figure*}

The core of LogicLens is a graph-based retrieval-augmented generation (GraphRAG) system. The agent constructs context for LLM queries by retrieving relevant subgraphs from the complete graph constructed in the previous phase. This approach enables the system to answer complex queries about the system, both at a technical level (e.g., specific code implementations) and at a functional level (e.g., workflows across multiple projects).

The agent reactively selects appropriate tools based on the user's query (as shown in Figure \ref{fig:react_agent}). After receiving results from a tool, it decides whether to gather additional information using other tools or generate a final response. This reactive mechanism allows the agent to handle diverse inquiries across multiple repositories efficiently.

The agent uses a suite of tools to extract information from the knowledge graph at varying levels of detail. Each tool targets a specific type of query, enabling precise retrieval of relevant nodes, subgraphs, or source code.

\paragraph{Projects Tool.} 
Returns a subgraph focused on specific projects, suitable for queries about a particular project’s structure or workflow. The process involves: (1) selecting \texttt{Project} nodes using semantic search, (2) filtering project nodes based on query relevance, (3) expanding the subgraph with adjacent code nodes, (4) filtering code nodes for query relevance, and (5) constructing the induced subgraph.

\paragraph{Entities Tool.} 
Returns a subgraph centered around domain entities. Domain entities are key concepts or objects in the application’s domain, such as \texttt{Order} in an order management system. The process includes: (1) selecting \texttt{Entity} nodes via semantic search, which retrieves nodes most relevant to the query based on their meaning rather than keywords, (2) filtering these entities using similarity scores between query embeddings and entity embeddings, (3) including incoming edges to selected entities, and (4) constructing an induced subgraph from the filtered nodes and edges. This tool is essential for functional queries that span multiple projects, tracking how domain entities propagate across systems.

\paragraph{Codes Tool.} 
Returns a subgraph focused on specific code elements, handling technical queries about implementations. The process is similar to the Entities Tool: (1) selecting \texttt{Code} nodes through semantic search, (2) filtering nodes for query relevance, and (3) constructing the induced subgraph from selected nodes and their relationships.

\paragraph{Graph Query Tool.} 
Executes arbitrary queries against the graph database to retrieve nodes, subgraphs, counts, or aggregated data. This tool allows direct access to structural or analytical information from the graph, enabling queries that go beyond predefined tool operations.

\paragraph{Source Tool.} 
Returns the source code of specific code nodes, used when the agent needs to examine full implementation details beyond the graph representation.

The choice between the \textbf{Projects Tool} and the \textbf{Entities Tool} depends on the type of query: project-focused questions use the Projects Tool for a localized view of related code, while cross-project functional questions use the Entities Tool to track workflows across repositories. By leveraging semantic triples tied to domain entities, the agent can retrieve relevant code across project boundaries, providing a comprehensive, system-wide perspective.

\section{Evaluation}

We evaluated LogicLens on a \textbf{real-world enterprise system} consisting of multiple repositories, microservices, and interdependent components.  
The goal was to measure the agent's ability to answer complex queries about system architecture, troubleshooting, and impact analysis.

Our evaluation approach is inspired by the \textbf{RACCCA framework} \cite{wadsworth2024frameworks}, adapted to a \textit{one-shot prompting} setup, where each question is processed independently without conversational memory. 
We employ a one-shot prompting setup, where each question is processed independently, to reflect realistic stateless deployment scenarios and ensure that evaluation metrics capture the model's ability to handle isolated queries without relying on accumulated conversational context.

\paragraph{Experimental Setup.}
To evaluate our system, we built a dataset of 30 representative questions in collaboration with experts in the field. These questions are designed to assess different aspects of the system comprehension, categorized into three groups:
\begin{description}
    \item[\textbf{Factual Questions} ] probe concrete information about system components. For example, \textit{``Which modules handle authentication?"}

    \item[\textbf{Multiple Source Linking Questions}] require synthesizing information across different parts of the codebase to explain complex workflows. An example is: \textit{``How does the account creation flow work?"}

    \item[\textbf{Predictive Questions}] challenge the system to go beyond factual retrieval by performing root cause analysis and inference. For instance: \textit{``Why am I getting an InvalidConfig error in the API project?"}
\end{description}

Each question was provided as a single prompt to simulate real-world usage scenarios. The agent generated a complete response, without follow-up prompts or interactive clarification, allowing us to assess its ability to provide comprehensive responses in a single interaction.

\paragraph{Evaluation Metrics.} A human expert annotated the responses 
using three qualitative metrics, each rated on a three-point scale:

\begin{description}
    \item[Accuracy:] Technical correctness of the information provided.
    \begin{itemize}
        \item \textit{High}: All factual claims are correct and verifiable 
        against the codebase; no technical errors or hallucinations.
        \item \textit{Medium}: Minor inaccuracies or ambiguities present, but core technical content is sound.
        \item \textit{Low}: Contains significant factual errors, 
        hallucinations, or misleading information.
    \end{itemize}
    
    \item[Completeness:] Coverage of all relevant aspects of the question.
    \begin{itemize}
        \item \textit{High}: Addresses all key aspects of the query; 
        provides sufficient detail for the question's needs.
        \item \textit{Medium}: Covers main points but omits some relevant 
        details or secondary aspects.
        \item \textit{Low}: Misses critical information or provides only 
        superficial coverage.
    \end{itemize}
    
    \item[Coherence:] Clarity, structure, and logical consistency 
    of the generated text.
    \begin{itemize}
        \item \textit{High}: Well-structured, easy to follow, with clear 
        explanations and logical flow.
        \item \textit{Medium}: Generally understandable but with some 
        organizational issues or unclear passages.
        \item \textit{Low}: Confusing structure, unclear explanations, 
        or logically inconsistent statements.
    \end{itemize}
\end{description}

The results over the 30 questions are reported in Table \ref{tab:results-baseline} for the baseline and in Table \ref{tab:results-logiclens} for the current iteration of LogicLens. 

As a baseline, we constructed a reactive agent using n8n\cite{n8n} with access to vector-based indexing of the entire codebase. Specifically, the baseline system chunked the source code and stored it in Qdrant\cite{qdrant}, a vector database, enabling semantic retrieval through embedding similarity. The agent could query this vector store to retrieve relevant code snippets in response to user questions.

The results over the 30 questions are reported in Table~\ref{tab:results-baseline} for the baseline and in Table~\ref{tab:results-logiclens} for the current iteration of LogicLens.

\begin{table}[t]
\centering
\begin{tabular}{lccc}
\toprule
\textbf{Metric} & \textbf{High (\%)} & \textbf{Medium (\%)} & \textbf{Low (\%)} \\
\midrule
Accuracy    & 0    & 65.2  & 26.0   \\
Completeness & 0    & 52.1  & 43.48  \\
Coherence    & 17.39 & 69.57 & 8.7    \\
\bottomrule
\end{tabular}
\caption{Baseline performance metrics across 30 evaluation questions. Values represent percentage of responses rated at each quality level by human expert annotation.}
\label{tab:results-baseline}
\end{table}

\begin{table}[t]
\centering
\begin{tabular}{lccc}
\toprule
\textbf{Metric} & \textbf{High (\%)} & \textbf{Medium (\%)} & \textbf{Low (\%)} \\
\midrule
Accuracy    & 69.5  & 30.5  & 0.0   \\
Completeness & 26.0  & 65.3  & 8.7   \\
Coherence    & 52.2  & 47.8  & 0.0   \\
\bottomrule
\end{tabular}
\caption{LogicLens performance metrics across 30 evaluation questions. Values represent percentage of responses rated at each quality level by human expert annotation.}
\label{tab:results-logiclens}
\end{table}

As shown in Tables~\ref{tab:results-baseline} and~\ref{tab:results-logiclens}, LogicLens demonstrates substantial improvements over the baseline system across all three evaluation metrics. 

The most significant improvement is observed in \textbf{Accuracy}, where LogicLens achieves 69.5\% high-accuracy responses compared to 0\% for the baseline. This represents a dramatic shift in the quality distribution: while the baseline system produced primarily medium-accuracy (65.2\%) and low-accuracy (26.0\%) responses, LogicLens eliminates low-accuracy responses entirely and shifts the majority into the high-accuracy category. This improvement can be attributed to the semantic graph representation, which enables more precise retrieval of relevant code elements and their relationships compared to pure vector similarity search.

\textbf{Coherence} shows a similar pattern of improvement, with LogicLens achieving 52.2\% high-coherence responses versus only 17.39\% for the baseline. The elimination of low-coherence responses (from 8.7\% to 0\%) indicates that the structured graph representation helps maintain logical consistency in the generated explanations. The baseline's reliance on isolated code chunks often resulted in fragmented or contextually disconnected responses, whereas LogicLens's ability to traverse semantic relationships produces more cohesive narratives.

\textbf{Completeness} remains the most challenging metric for both systems, though LogicLens shows meaningful progress. While neither system achieves a majority of high-completeness responses, LogicLens increases this category from 0\% to 26.0\% and shifts responses from low (43.48\% to 8.7\%) to medium (52.1\% to 65.3\%) completeness. 

The iterative improvements to LogicLens, including the introduction of specialized tools for cross-repository traversal and domain entity resolution, were specifically targeted at increasing completeness scores. However, the results suggest that achieving consistently high completeness remains an open challenge, likely requiring more sophisticated strategies for information aggregation and summarization across large codebases.

\section{Emergent Behavior.}

During the evaluation of LogicLens in real-world contexts, we observed several unexpected and valuable use cases that emerged naturally from the system's semantic graph structure. These \emph{emergent behaviors}—capabilities that arise from the interaction between the graph's design and user queries, rather than being explicitly programmed—extended beyond the system's primary objectives of facilitating onboarding and troubleshooting.

\paragraph{Impact Analysis.} 
By leveraging its understanding of existing workflows, LogicLens can analyze the impact of newly described features on existing flows, code, and projects. For example, when asked \textit{``Which projects would be impacted by introducing this new User type? What code interventions would be necessary?"}, the agent can trace through the semantic graph to identify all affected components and their interdependencies.

This analysis can be further refined by asking the agent to hypothesize specific code modifications that would be required to implement the proposed changes.

\paragraph{Log Fragment Debugging.}
By simply providing a log fragment containing errors from a project service, LogicLens can formulate multiple hypotheses about the root causes. The agent's analysis extends beyond the immediate code context surrounding the log entry to consider the broader workflows that led to that execution point, highlighting issues that would otherwise be difficult to identify through traditional debugging approaches.

\paragraph{Symptom-Based Debugging.}
Similar to log-based debugging, the generation process also works effectively when describing high-level perceived problems. For instance, when presented with a user complaint such as ``I cannot create an order---the screen shows OK but no order appears in the list,'' the agent reconstructs the complete workflow and identifies potential failure points that may have occurred throughout the process.

\paragraph{High-Level Architectural Views.}
Having comprehensive knowledge of workflows around domain entities, LogicLens can generate sequence diagrams of flows and high-level component schemas. This capability provides developers with valuable architectural insights and documentation that emerges naturally from the semantic graph structure.

\section{Conclusion}

In this work, we present \textit{LogicLens}, a novel system that combines a semantic multi-repository code graph with a reactive conversational agent to support troubleshooting and onboarding in complex software ecosystems. The proposed three-phase graph construction approach integrates structural code relationships with semantic and functional knowledge extracted via Large Language Models, resulting in a rich and multi-dimensional representation of the system.

The evaluation revealed several emergent behaviors, including impact analysis, symptom-based debugging, and high-level architectural insight generation. These capabilities go beyond initial design goals and highlight the effectiveness of combining graph-based knowledge representations with advanced language models to reason over large and complex codebases.

\paragraph{Future Work.}
The current capabilities of LogicLens can be extended in several directions. First, structural graph construction is currently language-dependent, as it relies on parsing the abstract syntax tree (AST) of the source code. Future work will focus on supporting additional programming languages through the development of new parsers and query templates, enabling LogicLens to operate on polyglot codebases. Second, we plan to enrich the graph by integrating additional software artifacts beyond the source code, such as configuration files, API specifications, and documentation. This extension would allow the system to capture a more comprehensive view of both architectural and functional knowledge. Moreover, incorporating build and deployment information, such as Docker configurations, CI/CD pipelines, and Kubernetes manifests, would enable the agent to reason about the software lifecycle, deployment topology, and runtime orchestration.

Overall, LogicLens represents a promising step towards context-rich developer assistants capable of bridging the gap between low-level code structure and high-level system understanding, enabling more effective navigation, comprehension, and reasoning across large-scale, heterogeneous software landscapes.








\bibliographystyle{named}
\bibliography{references}

\end{document}